\documentclass[epjCONF]{svjour}
\usepackage{graphics}
\usepackage[varg]{txfonts} 
\usepackage[latin1]{inputenc}
\usepackage{color}
\session-title{The 23rd International Conference on Atomic Physics ICAP 2012}
\begin{document}
\title{Accurate determination of the Boltzmann constant by Doppler spectroscopy: towards a new definition of the kelvin}
\author{Benoît Darquié\inst{1}\fnmsep\inst{2}\fnmsep\thanks{\email{benoit.darquie@univ-paris13.fr}} \and Sinda Mejri\inst{2}\fnmsep\inst{1} \and Papa Lat Tabara Sow\inst{1}\fnmsep\inst{2} \and Cyril Lemarchand\inst{2}\fnmsep\inst{1}\fnmsep\thanks{Present address: Université Paul Sabatier, Université de Toulouse, LCAR, 31062 Toulouse, France} \and Meriam Triki\inst{2}\fnmsep\inst{1} \and Sean K. Tokunaga\inst{2}\fnmsep\inst{1} \and Christian J. Bordé\inst{1}\fnmsep\inst{2}\fnmsep\inst{3} \and Christian Chardonnet\inst{1}\fnmsep\inst{2} \and Christophe Daussy\inst{2}\fnmsep\inst{1}}
\institute{CNRS, UMR 7538, LPL, F-93430 Villetaneuse, France \and Université Paris 13, Sorbonne Paris Cité, Laboratoire de Physique des Lasers, F-93430 Villetaneuse, France \and Laboratoire National de Métrologie et d'Essais-Système de Références Temps-Espace, UMR 8630 Observatoire de Paris, CNRS, UPMC, 75014 Paris, France}
\abstract{
Accurate molecular spectroscopy in the mid-infrared region allows precision measurements of fundamental constants. For instance, measuring the linewidth of an isolated Doppler-broadened absorption line of ammonia around 10 $\mu$m enables a determination of the Boltzmann constant $k_{\mathrm{B}}$. We report on our latest measurements. By fitting this lineshape to several models which include Dicke narrowing or speed-dependent collisional effects, we find that a determination of $k_{\mathrm{B}}$ with an uncertainty of a few ppm is reachable. This is comparable to the best current uncertainty obtained using acoustic methods and would make a significant contribution to any new value of $k_{\mathrm{B}}$ determined by the CODATA. Furthermore, having multiple independent measurements at these accuracies opens the possibility of defining the kelvin by fixing $k_{\mathrm{B}}$, an exciting prospect considering the upcoming redefinition of the International System of Units.
} 
\maketitle
\section{Introduction}
\label{sec:intro}

Atomic physics has brought us some of the most precise measurements of fundamental constants \cite{Chu1999,Biraben2011}, stringent constraints on their variations in time \cite{Bize2012}, clocks of the highest accuracy \cite{Rosenband2010}, quantum sensors \cite{Kasevich2007,Borde2002}, probes of general relativity \cite{Salomon2001} and physics beyond the standard model of particles and fields \cite{Amoretti2002}. This formidable discipline has historically gone hand in hand with the study of molecules, which contributed to the development of laser physics and high-resolution laser spectroscopy, showing for instance the first observation of the photon recoil splitting \cite{Hall1976}. Molecular systems provided the first natural-resonance-based frequency standard in 1949 using microwave transitions in ammonia \cite{Townes1951}. Optical frequency chains developed to measure molecular transition frequencies led to the accurate measurement of a fundamental constant, the speed of light \cite{Hall1972}. Note in passing that the need for higher resolution and frequency accuracy inspired the first ideas on the use of cold atoms and molecules \cite{Dychkov1989,Borde1994}. The potential for molecules to outperform precision measurements carried out on atoms, owing to their rich electronic, vibrational, rotational and hyperfine structure is becoming increasingly apparent. Molecules are now being (or have recently been) used to test fundamental symmetries such as parity \cite{Daussy1999,DeMille2008,Darquie2010} and time reversal (in \cite{Hinds2011}, the upper limit on the size of the electron electric dipole moment -- a signature of parity and time reversal if non-zero -- was set using YbF molecules, outperforming the previous limit set by measurements on atoms), to test the symmetrization postulate of quantum mechanics \cite{Tino2000}, to measure either absolute values of fundamental constants (electron-to-proton mass ratio \cite{Schiller2007} or Boltzmann constant as discussed in this paper), or to measure their variation in time (fine structure constant $\alpha$ \cite{Ye2006}, electron-to-proton mass ratio \cite{Amy-Klein2008,Levshakov2011}).

The current International System of Units (SI) is built on seven well-defined base units: the meter, the kilogram, the second, the kelvin, the ampere, the mole and the candela. The definition of some of these units is based on macroscopic objects or macroscopic properties of matter. Since 1889 the unit of mass has been defined by $\mathcal{K}$, the international prototype of the kilogram, a platinum cylinder located at Bureau International des Poids et Mesures (BIPM) in Sèvres (France). This definition has a major drawback. It is based on the mass of an artifact, whose composition inevitably fluctuates in time. In more than a hundred years, comparisons of $\mathcal{K}$ with other prototypes showed drifts on the order of 50~$\mu$g ($5\times10^{-8}$ in relative value). This directly impacts the definition of the mole, the ampere and the candela which depend on it. Just as unsatisfactory, the kelvin is defined by the temperature (273.16~K) of the triple point of water of a specific isotopic composition. The precision with which this isotopic composition can be achieved currently limits the reproducibility of the kelvin realization at a few $10^{-7}$. A revision of the SI is thus expected in the near future \cite{Borde2005,Mills2005,Mills2006,BIPM2010,BIPM2011}. One objective is to avoid using artifacts or macroscopic properties of matter. Another objective is to redefine units in terms of fundamental constants expected to not vary in time. A third objective is to derive base units from the time unit which is by far realized with the best accuracy. This was done in 1983 for the length unit. The velocity of light $c$ was fixed to $299~792~458$~$\mathrm{m}\cdot\mathrm{s}^{-1}$, defining thereby the metre, as the length travelled by light in vacuum in ~$1/299~792~458$~s. In the new SI four of the base units, namely the kilogram, the ampere, the mole and the kelvin, will be redefined, based on fixed numerical values of the Planck constant $h$, the elementary charge $e$, the Avogadro constant $N_{\mathrm{A}}$ and the Boltzmann constant $k_{\mathrm{B}}$, respectively. Let us detail the kelvin case. Temperature has a microscopic interpretation. It is proportional to the mean energy  per particle per degree of freedom $E=\frac{1}{2}k_{\mathrm{B}}T$. $h/E$ having the dimensions of time, fixing the value of $h$ and $k_{\mathrm{B}}$ is a way to define temperature by linking it to the time unit. The revision of the SI has been envisaged thanks to remarkable progress made over the last few years towards the experimental determination of $h$, $e$, $N_{\mathrm{A}}$ and $k_{\mathrm{B}}$.

We report on our latest progress on the determination of the Boltzmann constant $k_{\mathrm{B}}$ by accurate molecular spectroscopy in the mid-infrared region, using the Doppler Broadening Technique (DBT). The 2010 value recommended by the CODATA (Committee on Data for Sciences and Technology) is $k_{\mathrm{B}(2010)}=1.380~648~8(13)\times 10^{-23}$~J.K$^{-1}$, with an uncertainty of 0.9~ppm, based on eight published values \cite{Mohr2012}. Six of them were obtained using a single method, namely the Acoustic Gas Thermometry. We are aiming to provide a competitive measurement using an alternative method in order to contribute to the next CODATA value and to the new definition of the kelvin.

\section{The Doppler Broadening Technique}
\label{sec:DBT}

\subsection{Principle}
\label{sec:principle}

\begin{figure}
\resizebox{0.75\columnwidth}{!}{
  \includegraphics{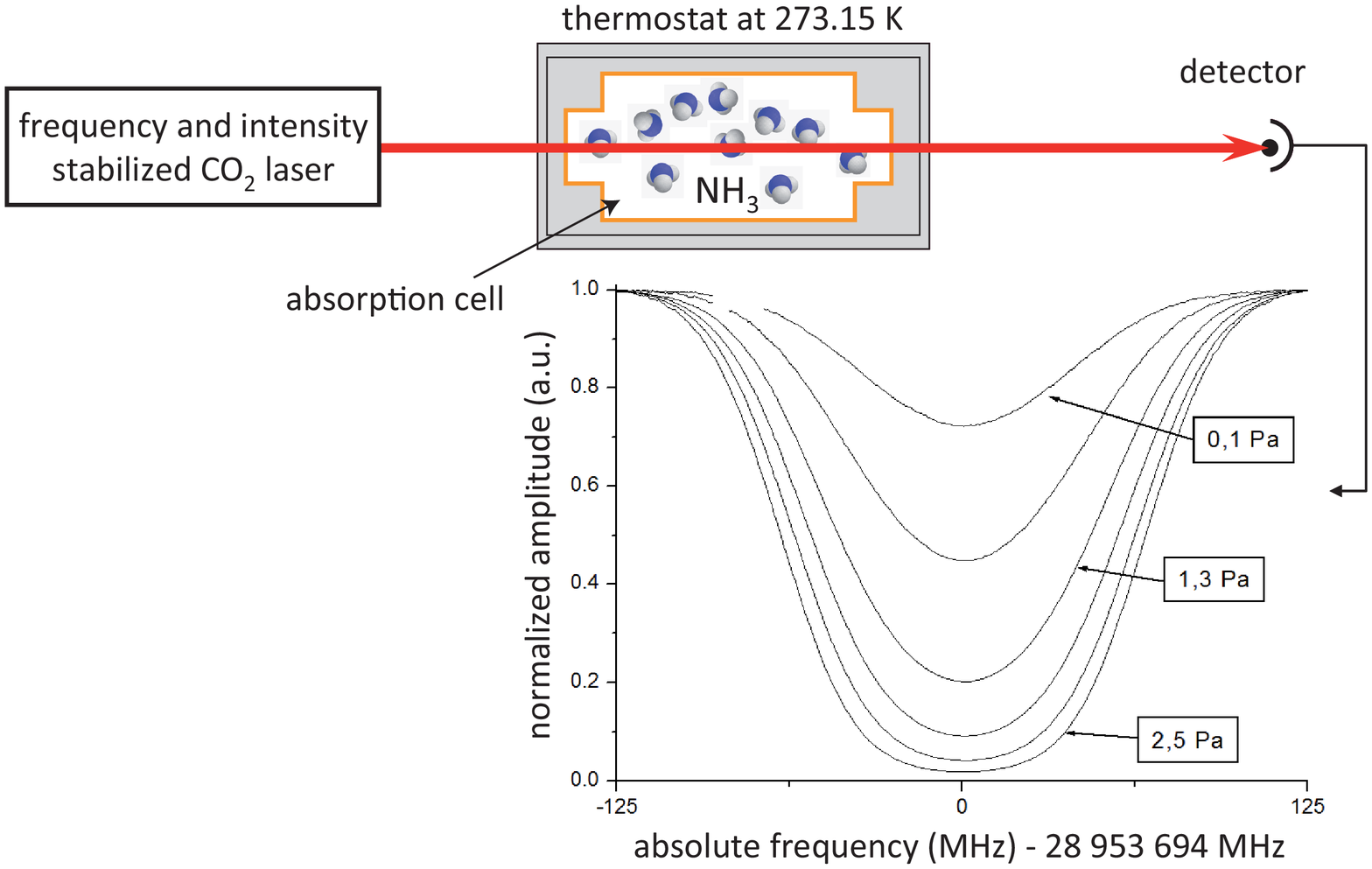} }
\caption{The Doppler Broadening Technique principle.}
\label{fig:DBT}
\end{figure}

The principle of the DBT \cite{Borde2002}, illustrated in Figure \ref{fig:DBT}, is to record the Doppler profile of an absorption line of a vapour in thermal equilibrium. In the Doppler regime, when all other broadening mechanisms are negligible, the recorded Gaussian profile derives from the Maxwell-Boltzmann distribution of velocities along the laser beam axis. The Boltzmamn constant is then calculated from $\Delta\nu_{\mathrm{D}}$ the half-Doppler width at $1/e$ :
$$
\frac{k_{\mathrm{B}}}{h}=\frac{m}{h}\frac{c^2}{T}\left(\frac{\Delta\nu_{\mathrm{D}}}{\nu_0}\right)^2
$$
where $m$ is the molecular mass, $\nu_0$ the central frequency of the molecular line and $T$ the gas temperature.

Using saturated absorption spectroscopy, we measure $\nu_0$ with a relative uncertainty of a few $10^{-9}$ \cite{Lemarchand2011}. The Watt balance experiment measures $h$ at the level of $5\times10^{-8}$ \cite{Mohr2012} and $h/m$ has an uncertainty of $10^{-8}$ deduced from atom interferometry experiments with rubidium \cite{Biraben2011} and atomic mass ratios measured in ion traps \cite{Pritchard1999}. The uncertainty on $k_{\mathrm{B}}$ is then dominated by uncertainties on the temperature and the Doppler width. The key ingredients required for an accurate determination of $k_{\mathrm{B}}$, are the following: \textit{(i)} control and accurate measurement in real time of the gas temperature, by placing the absorption cell in a thermostat, and \textit{(ii)} a precise record and reliable model of the lineshape in order to extract $\Delta\nu_{\mathrm{D}}$ from the data. In Section \ref{sec:T} we discuss how we are able to control and measure the temperature. Section \ref{sec:Doppler} details the Doppler width measurement and its accuracy.

\subsection{The experimental setup}
\label{sec:spectro}

The experimental setup has been described in detail in \cite{Djerroud2009,Lemarchand2010}. It contains a CO$_2$ laser stabilized on a saturated absorption line of osmium tetroxyde. Thanks to the resulting unique spectral properties (below 10~Hz width, 0.1~Hz frequency instability for 100~s integration time \cite{Bernard1997}), an incomparable control of the frequency scale is reached, a crucial element for a careful reproduction of the lineshape. The intensity of the laser beam is stabilized before being sent to the absorption cell filled with gaseous ammonia. NH$_3$ was chosen because of its strong $\nu_2$ absorption band which overlaps with the 8-12~$\mu$m spectral range of our ultra-stable spectrometer. Furthermore, this band exhibits well-isolated transitions which minimize any deformation of the line shape due to neighbouring lines. The absorption length of the cell can be adjusted from 37~cm in a single-pass configuration (SPC) to 3.5~m in a multipass configuration (MPC). By varying the pressure from 1 to 25 Pa (SPC) and 0.1 to 2.5~Pa (MPC), absorption amplitude can be varied between 20 to 98\%. The laser frequency is tuned close to the $\nu_2$ saQ(6,3)\footnote{$J=6$ and $K=3$ are, respectively, the quantum numbers associated with the total orbital angular momentum and its projection on the molecular symmetry axis.} line of $^{14}$NH$_3$ (of measured central frequency $\nu_0=28~953~693.9(1)$~MHz), and scanned via a tunable electro-optic modulator. The absorption cell is placed inside a thermostat at 273.15~K (see Section \ref{sec:T}). In order to avoid pressure induced modification of the lineshape (see Section \ref{sec:lineshape}) while preserving a good signal-to-noise ratio, spectra are recorded at pressures around 1~Pa in MPC (the SPC is used when pressure effects are to be studied). Typical spectra are displayed on Figure \ref{fig:DBT}. They are recorded in steps of 500~kHz with 30~ms of integration time per point, leading to a typical signal-to-noise ratio of 10$^3$.

\section{Temperature control}
\label{sec:T}

This experiment requires the molecular gas to be maintained at a constant and homogeneous temperature. A thermostat has been designed in close collaboration with thermometry experts from Laboratoire Commun de Métrologie, LNE-CNAM. The absorption cell is placed inside a copper thermal shield which is itself inside a stainless steel enclosure immersed in about 1~m$^3$ of an ice-water mixture which stabilizes the temperature close to 273.15 K. The thermal shield is only linked to its enclosure via a copper ring. The absorption cell is mechanically linked to the thermal shield via Teflon holders to ensure that heat transfer between them is predominantly radiative. This, in turn ensures that the heat transfer is spatially smoothed, thereby reducing residual temperature gradients, but also increases the thermal time constant of the system (to about 10~h), which smoothes temperature drifts over similar time scales. The gas temperature is measured with Hart capsule standard platinum resistance thermometers (cSPRTs) directly placed inside vacuum, in contact with the absorption cell. Those cSPRTs are calibrated at the triple point of water and at the gallium melting point, and are coupled to a resistance measuring bridge (Guildline Instruments Limited, Ref. 6675/A) calibrated against a resistance standard with a very low-temperature dependence. In this thermostat the temperature of the gas probed by the laser beam shows an overall accuracy of $\sim1$~ppm, deduced from a thorough analysis of the thermal stability and homogeneity of the global temperature control system \cite{Lemarchand2010}.

\section{Doppler width measurement}
\label{sec:Doppler}

The Doppler width is obtained by fitting recorded spectra (see Figure \ref{fig:DBT}) to the Beer-Lambert law (with an added baseline):
\begin{equation}
P(\nu)=[P_0+s(\nu-\nu_0)]\exp[-AI(\nu-\nu_0)]
\label{eq:Beer-Lambert}
\end{equation}
with $P_0$ the incident power, $s$ the slope of the baseline, $\nu$ the laser frequency, $A$ the integrated absorbance (proportional to the pressure and the absorption length) and $I(\nu-\nu_0)$ the normalized absorption profile that contains the Doppler width $\Delta\nu_{\mathrm{D}}$. The exact form of the profile depends on the assumption made for the type of collisions between molecules, as will be discussed in Section \ref{sec:lineshape}.

\subsection{Accuracy}
\label{sec:syst}

We have listed and investigated a number of systematic effects that may affect our determination of the Boltzmann constant. Systematic shifts in our measurement have to be measured or estimated in order to either cancel or correct for them. To build a proper error budget, it is also essential to estimate the uncertainty on each correction.

\subsubsection{The lineshape problem}
\label{sec:lineshape}

When the pressure goes to 0~Pa, the collision rate goes to 0. The absorption profile $I(\nu-\nu_0)$ (see Equation \ref{eq:Beer-Lambert}) is only Doppler broadened and is a simple Gaussian profile of half-width at $1/e$ $\Delta\nu_{\mathrm{D}}$, of about 50~MHz in our experimental conditions\footnote{typical natural width for rovibrational levels are negligible, of the order of a few Hz.}. In practice however, collisions induce both an additional broadening and a shift. The overall lineshape is then a Voigt profile (VP), \textit{i.e.} the convolution of the above mentioned Gaussian profile with a Lorentzian profile. In our experimental conditions the collisional broadening and shift (the actual half-width at half-maximum $\gamma$ and shift $\delta$ of the Lorentzian function) are typically 100~kHz/Pa and 1~kHz/Pa respectively. The VP is probably the most widely used model in the spectroscopy community. However more subtle effects cannot be ignored in our case. Collisions induce an auto-confinement of molecules by the surrounding molecules which leads to a narrowing of the Doppler contribution to the lineshape \cite{Dicke1953}. This narrowing is important when the pressure is such that this confinement is at the scale of the wavelength, the so-called Lamb-Dicke-Mössbauer (LDM) regime. In our case the LDM narrowing is expected to be about $\beta\sim10$~kHz/Pa \cite{Pine1993}, where $\beta$ is the frequency of velocity-changing collisions, an additional parameter of the lineshape. Depending on the assumption made for molecular collisions, adding this contribution to a VP leads to different lineshapes. The Galatry profile (GP) assumes soft collisions against light particles in a typical Brownian motion picture, in which the memory of the initial velocity is conserved after many collisions \cite{Galatry1961}. The Rautian profile (RP) assumes strong hard sphere collisions against infinitely massive particles, after each of which the memory of the initial velocity is lost \cite{Ghatak1964,Rautian1964}. Another subtle effect is the fact that the collisional broadening and shifting actually slightly depend on the molecular speed. Taking this into account leads to the speed-dependent Voigt profile (SDVP) with an overall narrowing compared to the usual Voigt profile. The impact of all collisional effects considered here are small in the typical low pressure experimental conditions favored for the determination of $k_{\mathrm{B}}$. In the general case, the lineshape can result from the combination of all these effects leading to more complex lineshapes \cite{Hartmann2008}.

\begin{figure}
\resizebox{0.75\columnwidth}{!}{
  \includegraphics{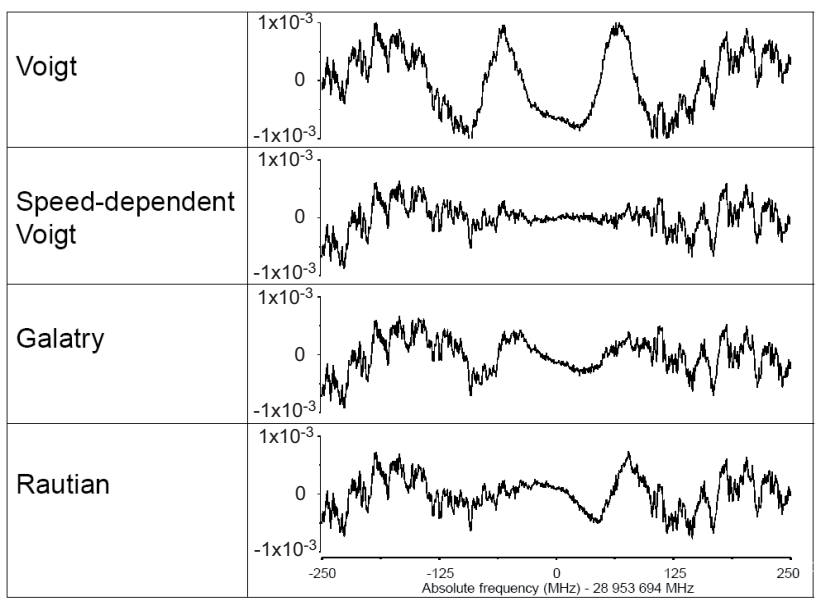} }
\caption{Normalized residuals for nonlinear least-squares fits of the same average of 24 spectra, recorded at 17.3~Pa, to the exponential of a Voigt, a speed-dependent Voigt, a Galatry and a Rautian profile.}
\label{fig:fit}
\end{figure}

Knowing the true lineshape of a given absorption line is fundamentally difficult. Until recently, the main source of uncertainty on determining $k_{\mathrm{B}}$ was the lack of knowledge of the correct collisional model for the gas, and therefore the correct lineshape to use for fitting to our experimental data. It for instance, constituted 97\% of the 144~ppm standard uncertainty published in \cite{Lemarchand2011}. The influence of the lineshape model on the spectroscopic determination of the Boltzmann constant has been raised by us and other groups working with O$_2$ and H$_2$O \cite{Lemarchand2011,Lemarchand2010,Triki2012,Ciurylo2010,Ciurylo2010b,Gianfrani2011,Ciurylo2011,Gianfrani2011b}. We recently carried out an extensive lineshape study at relatively high pressure in SPC, in order to enhance collisional effects and to determine the most suitable model for our data \cite{Triki2012}. The residuals obtained after fitting the average of 24 spectra recorded at 17.3~Pa to the exponential of a VP, SDVP, GP and RP are displayed in Figure \ref{fig:fit}. Residuals for the VP, GP, and RP exhibit structures with amplitudes respectively 10.3, 4.3 and 6.7 times larger than the rms noise level, clearly indicating that these profiles do not match the experimental data. By contrast, analysis with the SDVP demonstrates a good agreement between this model and the experimental lineshape. In the covered pressure range, the deviation from a Voigt profile of the $\nu_2$ saQ(6,3) line of $^{14}$NH$_3$ seems dominated by speed-dependent effects and not LDM effects. The following speed-dependent lineshape parameters have been obtained at the few percent level: $\gamma=120(3)$~kHz/Pa, $\delta=1.2(1)$~kHz/Pa, $m=0.360(9)$ and $n=-3.8(3)$\footnote{the $m$ and $n$ parameters are related to the speed-dependence of the collisional broadening and shift: under the hypothesis that the molecular interaction potential $V$ is proportional to some inverse power of the internuclear separation $R$, $V\sim1/R^q$, where $q$ is determined by the kind of the interaction, the dependence of the collisional width and shift versus the relative collisional speed $v_{\mathrm{r}}$ are given by $\Gamma(v_{\mathrm{r}})\sim v_{\mathrm{r}}^m$ and $\Delta(v_{\mathrm{r}})\sim v_{\mathrm{r}}^n$ respectively \cite{Triki2012}.}.

We have clearly identified the SDVP as the most suitable lineshape model. Having extracted the lineshape parameters $m$ and $n$ with good accuracy from data taken at high pressure, we fix these parameters when fitting the low pressure data used to determine $k_{\mathrm{B}}$. The uncertainty on $k_{\mathrm{B}}$ due to collisional effects is then only limited by our knowledge on the associated parameters $m$ and $n$. Simulated SDVP spectra corresponding to typical experimental conditions for $k_{\mathrm{B}}$ determination have been fitted taking into account uncertainties on $m$ and $n$ (see above) in order to estimate the impact of these uncertainties on the Doppler width accuracy. We followed the line-absorbance based analysis method recently proposed in \cite{Gianfrani2011c}, which leads to an estimated uncertainty of 0.9~ppm (respectively 1.8~ppm) on $\Delta\nu_{\mathrm{D}}$ (respectively $k_{\mathrm{B}}$), if spectra are recorded at ammonia pressures below 2~Pa.

\subsubsection{Hyperfine structure}
\label{sec:HFS}

The saQ(6,3) line has been chosen because it is a well-isolated rovibrational line with long-lived levels. However, owing to the non-zero spin values of the N and H nuclei, an unresolved hyperfine structure (HFS) is present in the Doppler profile. It causes a broadening of the lineshape according to the relative position and strength of the 78 individual hyperfine components. Although the 12 strongest components are spread about 50~kHz, weaker lines around $\pm\sim600$~kHz away from the main structure must also be considered, as they give the largest contribution to the HFS induced broadening. The relative increase in the linewidth scales as the square of the ratio $\Delta\nu_{\mathrm{hyp}}/\Delta\nu_{\mathrm{D}}$, with $\Delta\nu_{\mathrm{hyp}}$ being the spread of the HFS. For $\Delta\nu_{\mathrm{D}}\sim50$~MHz, the impact is a few ppm.

Accurate knowledge of the HFS is required in order to apply the right correction to the Doppler width. Ultra-high resolution saturated absorption spectra of the saQ(6,3) line were measured in a 3-m long Fabry-Perot cavity \cite{Lemarchand2011}. A combined analysis of our infrared spectra and microwave data from the literature led to an accurate determination of the hyperfine constants of the transition. The position of the 78 components and their relative intensities were then calculated, together with the associated uncertainties. In order to estimate this HFS-induced systematic effect, we fit to a single SDVP a simulated spectrum composed of the sum of 78 SDVP with positions and intensities corresponding to the precisely determined HFS. This leads to an overestimation of $\Delta\nu_{\mathrm{D}}$ of $4.356(13)$~ppm. The correction to be applied on the value of $k_{\mathrm{B}}$ is thus -8.71(3)~ppm, with negligible uncertainty.

\subsubsection{Other systematic effects}
\label{sec:otherSyst}

A number of other systematic effects have been considered and shown to affect the error budget of our measurement at a negligible level \cite{Lemarchand2011}. Laser-power-related systematic effects such as non-linearity in the detection setup, saturation of the probed transition and differential saturation of the different hyperfine components were simulated and investigated experimentally. A systematic shift results from amplitude modulation of the laser beam at 40~kHz used for noise filtering. It can be readily estimated from simulations with negligible uncertainty. Typical lifetimes for rovibrational levels are of the order of 100~ms, translating into a minute natural width. Our fitting procedure \cite{Triki2012} is such that presence of outgassing or partial pressure of impurities has rigourously no impact. The world record frequency stability and accuracy of our CO$_2$ laser (see Section \ref{sec:spectro}) lead to negligible contributions from the laser linewidth or the linearity and accuracy of the laser frequency scale.

It is sometimes mentioned that there should be an additional broadening of the lineshape due to the finite transit time of particles through the laser beam. It has recently been shown that this is strikingly not the case in linear absorption spectroscopy, provided that the medium is uniform and isotropic \cite{Borde2009}. Furthermore, this absence of transit time broadening holds true even after introducing a speed-dependent broadening and shift in the calculations \cite{Lemarchand2012}. All transit effects are already included in the inhomogeneous Doppler broadening, independently of the optical quality of the laser beam wavefronts and effective diameters. Rigourously no systematic effect due to the laser beam geometry is thus expected on the Doppler width measurement.

\begin{figure}
\resizebox{0.75\columnwidth}{!}{
  \includegraphics{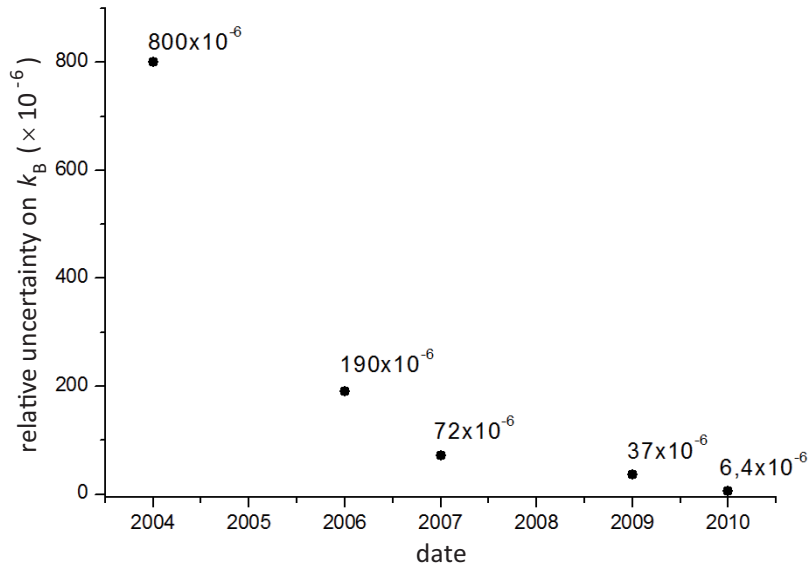} }
\caption{Improvement in the statistical uncertainty of $k_{\mathrm{B}}$ measured at Laboratoire de Physique des Lasers from 2004 onwards.}
\label{fig:stat}
\end{figure}

\subsection{Statistical uncertainty}
\label{sec:stat}

7171 spectra, corresponding to 70~h of integration time, were recorded for pressures ranging between 0.2 and 2.5~Pa. However, these were recorded in 2010, prior to the extensive lineshape study described in Section \ref{sec:lineshape}, under non-optimal experimental conditions. Regardless, the data was fitted with the exponential of a Galatry profile assuming the observed deviation from a Voigt profile to be a consequence of the LDM narrowing \cite{Lemarchand2011}. The purpose of this fitting exercise was two-fold. Firstly, the discrepancy between $\Delta\nu_{\mathrm{D}}$ values obtained by fitting Galatry and Voigt profiles estimates the systematic error which would occur from not knowing the true lineshape. We indeed find large systematic shift and a standard uncertainty of 144~ppm \cite{Lemarchand2011}, highlighting the importance of seeking the correct profile. Secondly, it demonstrates our ability to reach a statistical uncertainty below 10~ppm. The mean of the 7171 extracted Doppler widths was calculated to be $\Delta\nu_{\mathrm{D}}=49.885~90(16)$~MHz, leading to a statistical uncertainty on $k_{\mathrm{B}}$ of 6.4~ppm.

The improvement in the statistical uncertainty of $k_{\mathrm{B}}$ obtained at Laboratoire de Physique des Lasers from 2004 onwards \cite{Lemarchand2011,Djerroud2009,Lemarchand2010,Daussy2005,Daussy2007,Djerroud2007} is illustrated in Figure \ref{fig:stat}. To our knowledge, 6.4~ppm represents the best statistical uncertainty ever obtained for the measurement of $k_{\mathrm{B}}$ using the DBT.

Other groups followed our initiative in measuring $k_{\mathrm{B}}$ using the DBT, but probing other species such as CO$_2$, H$_2$O, C$_2$H$_2$ and Rb \cite{Gianfrani2011,Gianfrani2008,Shimizu2008,Luiten2011,Sun2011}. Table \ref{tab:1} gives the corresponding state-of-art for the statistical uncertainty obtained on $k_{\mathrm{B}}$. Our group is currently holding the record and leading the field.

\begin{table}
\caption{International state-of-the-art for the statistical uncertainty (except for Hu \textit{et al.} for which only the global standard uncertainty can be found in the literature) obtained on $k_{\mathrm{B}}$ using the DBT.}
\label{tab:1}
\begin{tabular}{lllll}
\hline\noalign{\smallskip}
place & group & species & wavelength ($\mu$ m) & fractional uncertainty (ppm) \\
\noalign{\smallskip}\hline\noalign{\smallskip}
USTC, China & S.M. Hu \textit{et al.} \cite{Sun2011} & $^{12}$C$_2$H$_2$ & 0.8 & 2000 \\
NMIJ, Japan & T. Shimizu \textit{et al.} \cite{Shimizu2008} & $^{13}$C$_2$H$_2$ & 1.5 & 1200 \\
UWA, Australia & A. Luiten \textit{et al.} \cite{Luiten2011} & $^{85}$Rb & 0.8 & 398 \\
UniNa2, Italy & L. Gianfrani \textit{et al.} \cite{Gianfrani2008} & $^{12}$C$^{16}$O$_2$ & 2 & 90 \\
UniNa2, Italy & L. Gianfrani \textit{et al.} \cite{Gianfrani2011} & H$_2$$^{18}$O & 1.4 & 80 \\
UP13, France & our group \cite{Lemarchand2011} & $^{14}$NH$_3$ & 10.3 & 6.4 \\
\noalign{\smallskip}\hline
\end{tabular}
\end{table}

\section{Conclusion}
\label{sec:conclusion}

\begin{figure}
\resizebox{0.75\columnwidth}{!}{
  \includegraphics{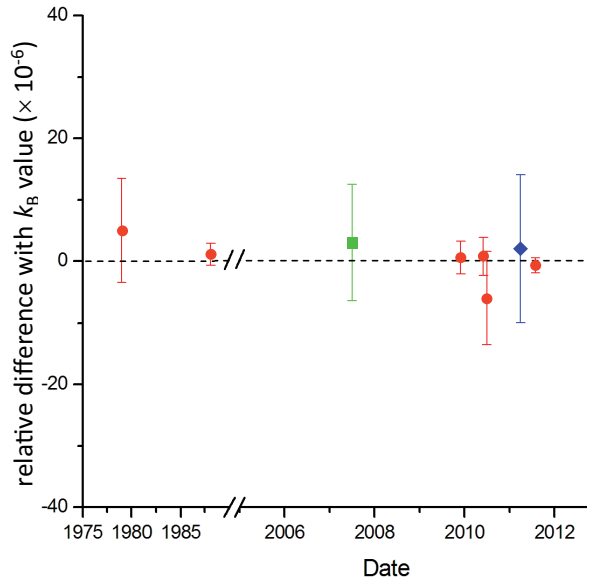} }
\caption{Comparison between the eight measurements accepted to contribute to the current CODATA value of $k_{\mathrm{B}}$ \cite{Mohr2012} (\textcolor{red}{{\large $\bullet$}}: Acoustic Gas Thermometry, \textcolor{green}{{\footnotesize $\blacksquare$}}: Refractive Index Gas Thermometry, \textcolor{blue}{{\footnotesize $\blacklozenge$}}: Johnson Noise Thermometry). Vertical axis: relative difference between measurement and the current CODATA value of $k_{\mathrm{B}}$. Our next measurement using the Doppler Broadening Technique will have a sub-10~ppm competitive uncertainty.}
\label{fig:conclusion}
\end{figure}

Taking into account all systematic effects, including temperature control, leads to an overall correction to be applied to $k_{\mathrm{B}}$ with an uncertainty on this correction at the level of a few ppm. This is valid provided that data is recorded under the optimized experimental conditions determined by these studies and provided that spectra are fitted to the SDVP, identified as the most suitable lineshape for our measurements. We are able to reach a statistical uncertainty on $k_{\mathrm{B}}$ of 6.4~ppm after 70~h of acquisition, as demonstrated during a previous data run. Our results are considered by the \textit{2010 CODATA Recommended Values of the Fundamental Physical Constants} \cite{Mohr2012} to be "of interest" because they are "obtained from a relatively new method that could yield a value with a competitive uncertainty in the future". To conclude, the first determination of $k_{\mathrm{B}}$ using the DBT with a combined uncertainty below 10~ppm is well within reach. A new series of data will soon be recorded.

The eight measurements accepted to contribute to the current CODATA value of $k_{\mathrm{B}}$  are displayed in Figure \ref{fig:conclusion} \cite{Mohr2012}. None of them use the DBT. Moreover, six of them were obtained using Acoustic Gas Thermometry (AGT). Refractive Index Gas Thermometry (RIGT) and Johnson Noise Thermometry (JNT) led to the two other measurements. JNT is based on the temperature dependence of the mean square noise voltage, developed in a resistor by thermal motion of the charge carriers. The principle of the gas-based measurements, AGT and RIGT, is to determine the gas-particle's microscopic energy, $k_{\mathrm{B}}T$, by measuring a macroscopic temperature-dependent gas property at known pressures: the speed of sound for AGT or the refractive index for RIGT. In comparison, the DBT has the advantage of being a direct measurement of the microscopic thermal energy. Moreover, the probed atoms or molecules belong to a single quantum level of a well-defined isotopic species, which avoids uncertainties coming from macroscopic quantities, an issue for other techniques. We would also like to stress the broad and generic nature of the DBT, as it can be used with any molecular as well as atomic (such as $^{85}$Rb in Australia) species (see Table \ref{tab:1}). It is worth mentioning that six of the eight above mentioned values were obtained between 2006 and 2010, illustrating the efforts made by several metrology institutes in preparation for the upcoming redefinition of the kelvin.

A measurement using the DBT at the level of or below 10~ppm will provide another value for $k_{\mathrm{B}}$ with competitive uncertainty, using an independent method. It should contribute to the next CODATA value and to the new definition of the kelvin.

\begin{acknowledgement}

This work is funded by CNRS, Laboratoire National de Métrologie et d'Essais and by the European Community (EraNet/IMERA). Authors would like to thank Y. Hermier, F. Sparasci, and L. Pitre from Laboratoire Commun de Métrologie LNE-CNAM for platinum resistance thermometer calibrations, and S. Briaudeau from Laboratoire Commun de Métrologie LNE-CNAM for helping in the design and setting up of the thermostat and temperature control devices.

\end{acknowledgement}

\end{document}